\documentclass[conference]{IEEEtran}
\IEEEoverridecommandlockouts
\usepackage{cite}
\usepackage[table]{xcolor}
\definecolor{lightgraycell}{gray}{0.9}
\usepackage{tcolorbox}
\usepackage{enumitem}
\usepackage{latexsym}
\usepackage{pifont}
\usepackage{amsmath}
\usepackage{amssymb}
\usepackage{amsfonts}
\usepackage{graphicx}
\usepackage{textcomp}
\usepackage{enumitem}
\usepackage{multirow}
\usepackage{soul}
\usepackage{tikz}
\usepackage{booktabs}
\usepackage{algorithm}
\usepackage{algpseudocode}
\usepackage{colortbl}
\usepackage{bm}
\usepackage{float}
\usepackage{filecontents}
\usepackage{CJK}
\usepackage{makecell}
\newcommand{\ourattack}{Invisible Hands}
\newcommand{\ourmethod}{Gradient-Data-Free-BFA}
\newcommand{\ourmethodabbr}{GDF-BFA}
\newcommand{\xmark}{\ding{55}}%

\newcommand{\Lc}{\mathcal{L}}
\newcommand{\Sc}{\mathcal{S}}

\newcommand{\Wc}{\mathcal{W}}

\newcommand\cbf{\mathbf{c}}

\newcommand\sbf{\mathbf{s}}
\usepackage{placeins}
\usepackage[hidelinks]{hyperref}

\begin{document}

\title{Invisible Hands: Gray-Box Bit Flip Attack for Steering LLMs Without Knowledge of Gradients, Data, and Weights}

\author{
\IEEEauthorblockN{
Abeer Matar A. Almalky$^{*,1}$,
Ziyan Wang$^{*,2}$,
Mohaiminul Al Nahian$^{*,1}$,
Li Yang$^{2}$,
Adnan Siraj Rakin$^{1}$
}
\IEEEauthorblockA{
\textit{$^{1}$Binghamton University \quad
$^{2}$UNC Charlotte}
}
\IEEEauthorblockA{
\{aalmalky, malnahian, arakin\}@binghamton.edu \\
\{zwang53, lyang50\}@charlotte.edu \\
* Equal contribution
}
}

\maketitle
\begin{abstract}
In recent years, large language models (LLMs) have achieved remarkable advances and are increasingly deployed in critical applications across diverse domains. This growing adoption raises urgent concerns about their security and robustness. In this work, we investigate the impact of Bit Flip Attacks (BFAs) on LLMs, which exploit hardware faults to corrupt model parameters, thereby threatening model integrity and performance. Existing BFA studies primarily assume a white-box setting with access to exact model weights and part of the dataset, and rely on progressive gradient-based bit-search strategies to identify vulnerable bits in model weights. However, gradient computation for LLMs is computationally expensive and memory intensive. In addition, assuming access to exact victim model weights and datasets is challenging due to increasingly strict user privacy regulations.
To address these challenges, we propose the first gray-box BFA framework for LLMs, \textit{\ourattack}, designed for efficient and practical deployment. Our method, \textit{\ourmethod}, identifies vulnerable weight bits without requiring knowledge of model weights, gradients, or sample data. It introduces novel vulnerability index metrics that estimate the weights of susceptibility based solely on model architecture (Grey-Box). By eliminating data access and gradient computation, our approach significantly reduces memory overhead and scales efficiently across tasks with constant complexity. Experiments on six open-source LLMs demonstrate that adversarial objectives can be achieved with minimal weight perturbations, highlighting the effectiveness and practicality of \ourattack.
\end{abstract}

\section{Introduction}
The continual advancement of Large Language Models (LLMs) has profoundly influenced the evolution of machine learning, driving the development of diverse applications and promoting their deployment in critical domains such as healthcare and economic systems \cite{kaddour2023challenges, chang2024survey}. As LLMs continue to expand in scale and societal influence, ensuring their robustness, safety, and security has become a central research imperative \cite{das2025securitychang2024survey}. Despite extensive attention to input-based threats in LLMs \cite{wang2025unique}, weight corruptions through hardware-based faults remain comparatively underexplored. Bit Flip Attack (BFA) \cite{yao2020deephammer, dong2023one} constitutes a particularly insidious category of hardware-level vulnerabilities that exploit remote side channels, such as rowhammer \cite{kim2014flipping}, to induce bit flips in DRAM. These perturbations can alter the memory regions that store model parameters in a resource-sharing environment, compromising the integrity, reliability, and overall trustworthiness of the model.

\begin{table}[t!]
\centering
\small
\setlength{\tabcolsep}{3.5pt}

\caption{Comparison of SOTA BFAs on LLMs: required knowledge, memory cost, and attack computational complexity.}

\begin{tabular}{|c|c|c|c|c|c|}
\hline
\textbf{\makecell{BFA \\on LLM}} &
\textbf{Gradients} &
\textbf{Weights} &
\textbf{Data} &
\textbf{\makecell{Memory\\Cost}} &
\textbf{\makecell{Attack \\Comp.}} \\
\hline

GenBFA \cite{das2024genbfa}
& $\checkmark$ & $\checkmark$ & $\checkmark$
& High 
& $O(N)$ \\
\hline

SBFA \cite{guo2025sbfa}
& $\checkmark$ & $\checkmark$ & $\checkmark$
& High
& $O(N)$ \\
\hline

SilentStriker \cite{xu2025silentstriker}
& $\checkmark$ & $\checkmark$ & $\checkmark$
& High
& $O(N)$\\
\hline

FlipLLM \cite{khalil2025flipllm}
& $\checkmark$ & $\checkmark$ & $\checkmark$
& High
& $O(N)$\\
\hline

\cellcolor{lightgraycell}\textbf{\textit{\makecell{\ourattack\\(ours)}}}
& \cellcolor{lightgraycell}\textbf{\xmark}
& \cellcolor{lightgraycell}\textbf{\xmark}
& \cellcolor{lightgraycell}\textbf{\xmark}
& \textbf{\makecell{Lower \\(8--10$\times$)}}
& \textbf{$O(1)$}\\
\hline

\end{tabular}

\label{tab:previous_works}
\vspace{-15pt}
\end{table}

Taking into account the application of LLMs in cloud services~\cite{chu2025cloud} and Machine Learning (ML) as a service platforms~\cite{raza2025industrial,LLMo}, the attack landscape of BFA in LLMs~\cite{das2024genbfa, guo2025sbfa, xu2025silentstriker, khalil2025flipllm} is increasingly becoming concerning. We have summarized the SOTA BFAs in LLMs in Table~\ref {tab:previous_works}, which highlights two fundamental limitations of them. First, prior works require computing the gradient of the weights, which is particularly expensive in the context of LLMs, thereby increasing the computational and memory costs of the attack algorithm. As shown in Table~\ref{tab:grad_mem}, computing gradients for the LLaMa-3-8B model on GPUs will require 68 GB of on-chip memory. Previous studies \cite{malladi2023fine, muhamed2406grass} have shown that backpropagation can require up to an order of magnitude more memory than inference, in some cases resulting in a 12× increase. This substantial memory overhead arises from the need to cache activations, parameters, and gradients. As summarized in Table~\ref{tab:grad_mem}, the memory required for gradient computation grows dramatically with the size of the model, a key bottleneck in training LLMs on large datasets often beyond the capabilities of a single high-end GPU (e.g., NVIDIA B200). Similar to the constraints of training in a low-resource environment, this problem is further exacerbated from an attacker's perspective, as the amount of computing and memory power also limits an attacker's capabilities.

A second limitation of existing BFAs in LLMs is their reliance on detailed knowledge of the victim model, including access to model weights and data. Early studies on adversarial weight corruption~\cite{rakin2019bit,rakin2020tbt} commonly assumed access to model parameters and samples from the victim dataset. As a result, these attacks effectively operate under a white-box setting. As summarized in Table~\ref{tab:previous_works}, current BFAs targeting LLMs follow the same assumption. They require direct access to model weights and either application-specific knowledge~\cite{xu2025silentstriker} or batches drawn from the training/ testing distribution~\cite{xu2025silentstriker, guo2025sbfa, khalil2025flipllm}. However, modern LLM deployments place a strong emphasis on protecting user data privacy in ML services and cloud platforms~\cite{li2024llm}. Consequently, such assumptions are often challenging, especially in privacy-sensitive domains such as healthcare. In addition, all existing BFAs are data-dependent, which leads to the lack of scalability across tasks. An attacker has to re-run the attack each time to target the same LLM architecture on a different application task or dataset. As shown in Table~\ref{tab:previous_works}, to attack an LLM for \textit{N} different tasks, the attacker has to run the attack algorithm \textit{N times}~\cite{xu2025silentstriker,guo2025sbfa,das2024genbfa,khalil2025flipllm}. 

To overcome these two limitations , we propose the first gray-box BFA in LLMs, \textbf{\ourattack}. As highlighted in Table~\ref{tab:previous_works}, \ourattack \space does not require knowledge about the victim model weights, gradients, or dataset/application domain; \textit{therefore \ourattack \space operates under a gray-box threat model}. Additionally, our attack is data-agnostic that only needs to be performed once to target an LLM across $N$ different applications/tasks, further reducing the attack complexity. 

\begin{table}[t]
\caption{Estimated total memory consumption for gradient computation under mixed-precision setting. 
Assumptions: batch-size = 16, sequence-length = 1024}
\scalebox{0.95}{
\begin{tabular}{|c|c|c|c|c|c|}
\hline
Model & 
\begin{tabular}[c]{@{}c@{}}Param Mem\\ (GB)\end{tabular} & 
\begin{tabular}[c]{@{}c@{}}Grad Mem\\ (GB)\end{tabular}  & 
\begin{tabular}[c]{@{}c@{}}Activation Mem\\ (GB)\end{tabular} & 
\begin{tabular}[c]{@{}c@{}}Total\\ (GB)\end{tabular} \\ \hline
LLaMA-3-8B      & 16  & 32  & 20 & $\approx$68 \\ \hline
LLaMA-2-13B     & 26  & 52   & 30 & $\approx$108 \\ \hline
\end{tabular}}
\label{tab:grad_mem}
\vspace{-2em}
\end{table}

However, designing a gray-box and data-agnostic BFA introduces two key technical challenges. First, the attacker lacks access to the gradient, which is fundamental to any adversarial attack. Whether an attacker wants to manipulate inputs or weights, the gradient dictates how much change is needed to leave an antagonistic impact on any ML workload. Second, the attacker has no access/knowledge of the application domain or data. Without data or domain knowledge, an attacker is almost unable to establish a connection between model weights and their adversarial outputs. \emph{In summary, without gradients and data, any adversarial attack is clueless in answering these questions: which parameter to attack? How much to change? And in which direction to change?}

To address these challenges, we propose a novel vulnerable bit search method \textit{\ourmethod \space (GDF-BFA)}, which consists of two components: a \emph{Layer Vulnerability Index (LVI)} to locate vulnerable layers of a specific foundation LLM and a gradient-free \emph{Weight Vulnerability Index (WVI)} to identify vulnerable weights in the identified layer, independent of any data or tasks. Since neither of these indices is data- or task- dependent, an attacker can compute them using any publicly available data, and that the sequence of vulnerable locations will remain the same even after fine-tuning the model for a downstream task.

\emph{\textbf{Our extensive evaluation across six different open-source LLMs shows that, regardless of the LLM task/application, an attacker can consistently achieve a targeted adversarial goal by targeting the exact same sequence of vulnerable locations of a model, which was identified using a public dataset and our proposed vulnerability metrics.}} Our results provide key insights into LLMs' vulnerability, showing that these foundation models have an in-built vulnerability that does not depend on user fine-tuning to different application tasks (classification or generative) or datasets. In the context of adversarial attack literature (input or weight attacks), \ourattack \space is the first to achieve this \textbf{BOLD} feat: successfully attacking LLMs in a gray-box setting: without requiring to compute gradients or knowledge of weights, domain/application, or the specific dataset.

\begin{figure*}[t]  
    \centering
    \includegraphics[width=0.85\textwidth,]{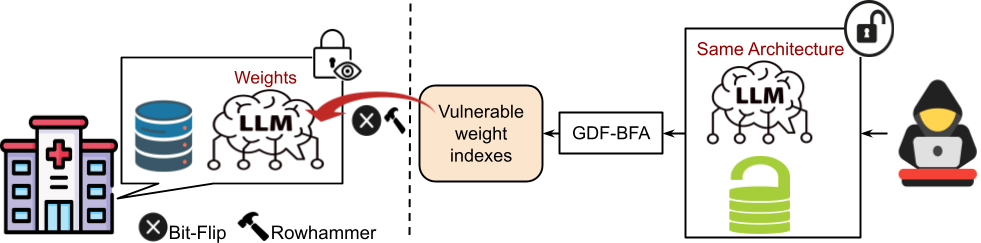}
    \caption{Threat Model. \textit{Left}: The target deployment environment (e.g., a hospital) protects model weights and datasets, while the model architecture remains publicly known. \textit{Right}: The attacker uses a publicly available LLM with the same architecture and any public dataset to identify the sequence of vulnerable weight indices offline using (\ourmethodabbr). The attacker then exploits hardware side-channel (e.g., Rowhammer) to flip bits at the identified locations to corrupt the deployed model.}
    \label{fig:threat-model}
    \vspace{-1em} 
\end{figure*}

\section{Background}
\subsection{Large Language Models (LLMs)}

LLMs are generally composed of embedding layers, multi-head attention layers, and feedforward (MLP) layers, which together enable the modeling of complex linguistic and semantic dependencies \cite{NIPS2017_3f5ee243}. Text generation in LLMs is based on predicting the probability distribution of the next token conditioned on the preceding context. By utilizing knowledge acquired from large-scale datasets, the model sequentially selects or samples tokens, allowing it to produce natural language text that is coherent and contextually appropriate.

\subsection{Bit-Flip Attack (BFA)}
BFA~\cite{rakin2019bit} is an adversarial technique that compromises models by selectively flipping critical bits in the binary representation of their weights. Through carefully chosen bit manipulations, BFA enables an attacker to pursue two adversarial objectives: untargeted and targeted. \textbf{Untargeted Attack.} In this setting, the adversary aims to broadly degrade the performance of the model. For example, in classification tasks, the goal is to flip a small number of highly sensitive bits that can cause the model to misclassify prompts into arbitrary classes. This objective can be expressed as follows:
\begin{equation}\label{eq:untargeted-attack}
    \max_{\hat{\Wc}} \ \mathbb{E}_{\sbf \sim \Sc}\left[ \Lc(F(\sbf, \hat{\Wc}), \cbf) \right]
\end{equation}
where $\Sc$ denotes the set of input prompts, $\sbf$ is an individual prompt, $\hat{\Wc}$ weights after flipping, and $\cbf$ is its corresponding ground-truth class. By maximizing the loss, the attacker forces the model to make incorrect predictions across the input space.

\textbf{Targeted Attack.} In this setting, the attacker  seeks to induce a specific malicious behavior; for example, forcing the model to consistently predict a predefined target class:
\begin{equation}\label{eq:targeted-attack}
    \min_{\hat{\Wc}} \ \mathbb{E}_{\sbf \sim \Sc}\left[ \Lc(F(\sbf, \hat{\Wc}), t) \right]
\end{equation}
where $\sbf$ denotes an input prompt from $\Sc$, and $t$ is the attacker-chosen target class. The goal is to minimize the loss with respect to this target class so that all prompts are misclassified into $t$. To achieve above mentioned objectives, \textit{the attacker is assumed to have access to model weights and training/test dataset}, which used in the evaluation of bit-level vulnerabilities. BFA is done through the following steps:

\textbf{\textit{First (in-layer search):}} The attacker computes weight gradients within each layer using the dataset. Bit flips are then simulated on the weights with the highest gradient magnitudes. The attacker evaluates the resulting degradation or targeted behavior and records the weight index that produces the greatest effect as the candidate flip for that layer.

\textbf{\textit{Second (intra-layer search):}} The attacker compares the candidate flips across all layers and selects the one that yields the highest attack impact as a winner candidate.

\textbf{\textit{Third (flipping permanently):}} The attacker repeats the above two steps iteratively until the attack goal is achieved or a maximum number of iterations is reached. These steps are performed on \textit{an offline copy of the model, and the attacker records the most effective flip locations.} Later, these recorded locations can be exploited using hardware-level side channels, such as RowHammer~\cite{mutlu2019rowhammer,197231,zhang2020pthammer,lin2025gpuhammer,244042,xiang2020open}, to induce the flipping.

\subsection{Bit-Flip attack on LLMs} 
Recent studies \cite{das2024genbfa, guo2025sbfa, coalson2024prisonbreak, xu2025silentstriker, khalil2025flipllm} have investigated the application of BFA on LLMs. However, these works largely adopt a white-box threat model, assuming that the attacker has access to the exact victim model weights and the training dataset or application domain. Furthermore, as summarized in Table~\ref{tab:previous_works}, existing approaches rely on gradient computations to evaluate the sensitivity of model layers or weights when identifying vulnerable bits for attack. As LLMs scale to billions of parameters, computing gradients over such models becomes computationally expensive and increasingly impractical. Additionally, given the growing emphasis on user data privacy and restricted data access in modern LLM deployments~\cite{li2024llm}, assuming dataset availability is often unrealistic. In contrast, our proposed method eliminates the need for accessing weights, gradients, and dataset or application domain, enabling a gray-box BFA framework that is both practical and computationally efficient for LLMs.

\section{Threat Model}
Nowadays, it is common practice for organizations that deploy LLMs in production to provide information through model cards, system documentation, or public announcements that describe model design, capabilities, and performance characteristics \cite{anthropic2025system, trokhymovych2024openmultilingualscoringreadability}. Consequently, obtaining architectural information for deployed models typically does not require unauthorized access or model extraction.

In contrast, the model parameters (weights) are treated as proprietary assets. Training large-scale models requires substantial computational resources and financial investment, often totaling millions of dollars in computation alone \cite{bommasani2022foundation, Sevilla_2022}. As a result, organizations restrict access to trained weights to protect intellectual property and mitigate potential misuse. Similarly, datasets used for training or fine-tuning are frequently protected due to privacy, regulatory, or commercial constraints, particularly in sensitive domains such as healthcare and finance \cite{protection2018general}. Accordingly, both model weights and training/testing data are typically assumed to be secure and inaccessible to adversaries.

Given these practical deployment realities, \ourattack \space operates under a practical threat model in which the attacker \textit{only } has access to the model architecture, which is publicly available. As illustrated in Figure~\ref{fig:threat-model}, the attacker leverages any publicly available model implementing the same architecture rather than the victim model itself. Importantly, the attack does not require access to the exact weights of the deployed model.

Furthermore, as in Figure \ref{fig:threat-model}, the adversary utilizes a publicly available dataset, even when it originates from a domain different from the victim task. Using an open-source model with the same architecture and publicly accessible data, the attacker performs offline analysis to identify the sequence of vulnerable weight indices that induce maximal performance degradation when corrupted. These identified locations can later be precisely flipped during deployment via conventional side-channel mechanisms to corrupt the deployed model during inference~\cite{gruss2016rowhammer,seaborn2015exploiting,van2016drammer,197231,zhang2020pthammer,xiang2020open,das2024genbfa, guo2025sbfa, coalson2024prisonbreak, xu2025silentstriker}. To achieve targeted bit-flip at the precise location an attacker does not need to know the exact weight values rather the storage location of the weights in memory which can be reversed engineered using existing attacks~\cite{gruss2018another, kwong2020rambleed}. This design enables the attack to generalize across multiple fine-tuned models that share the same underlying architecture.

\begin{figure*}[t]  
    \centering
    \includegraphics[width=\textwidth,]{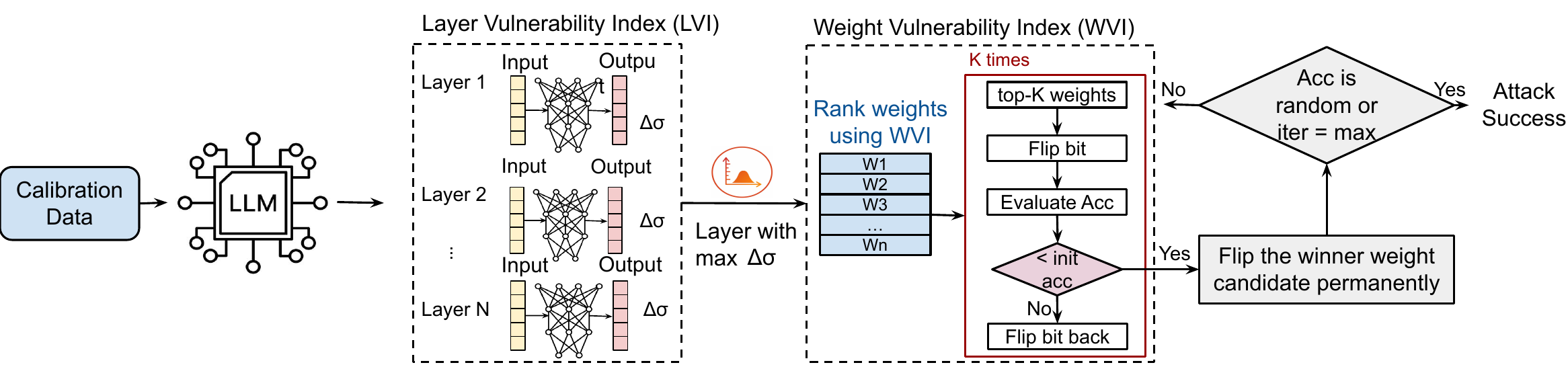}
    \caption{Overview of Gradient-Data-Free BFA (GDF-BFA). The attacker performs all computations offline using a copy of model with the same architecture as the victim model to identify and record the sequence of vulnerable weight locations. Later, during the online phase, the attacker flips the bits at these precomputed locations.}
    \label{fig:method}
    \vspace{-1em} 
\end{figure*}

\begin{figure}[t]  
    \centering
    \includegraphics[width=0.9\columnwidth]{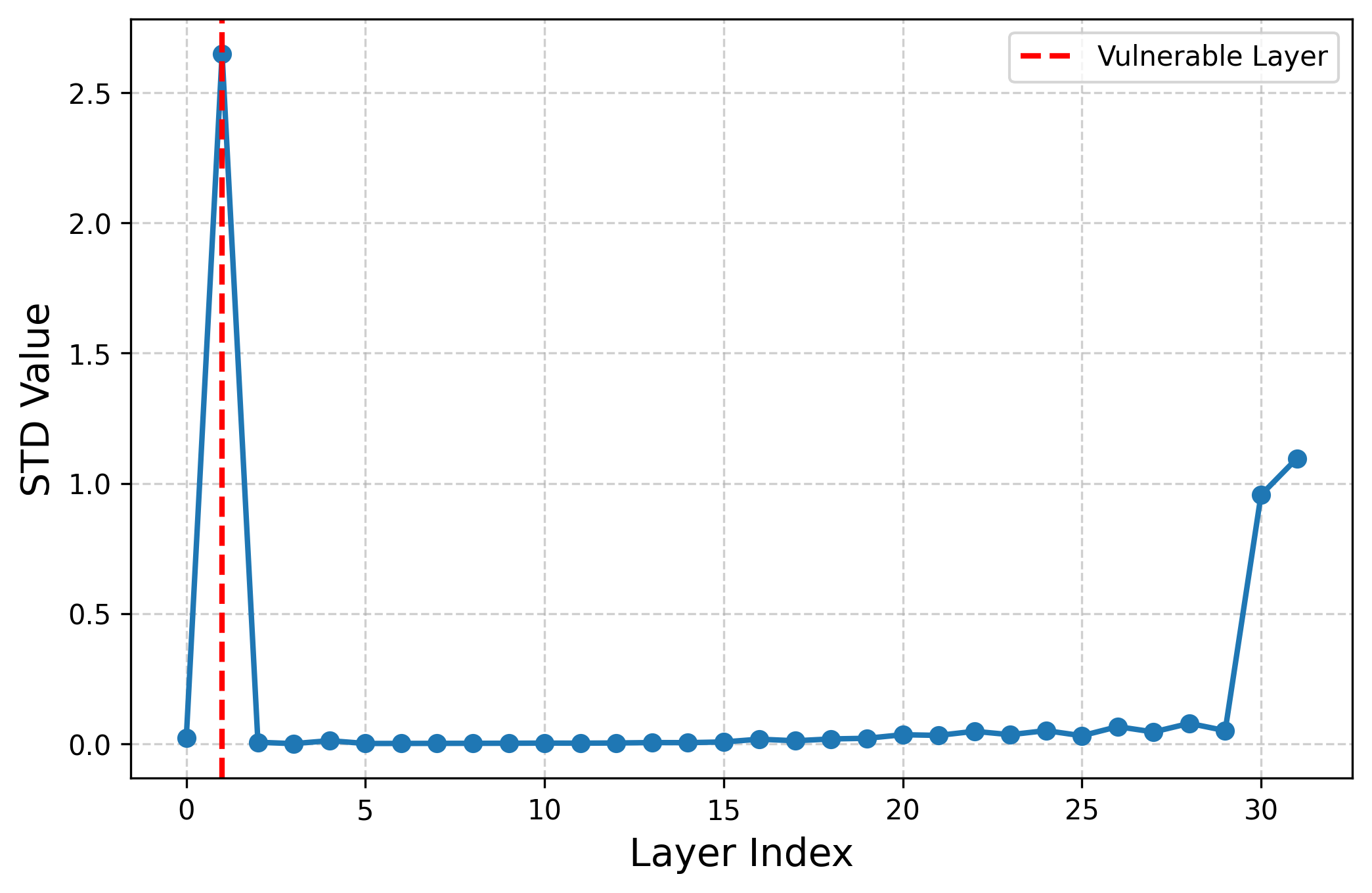}  
    \caption{LVI per layer with vulnerable layer marked in red for LLama-2-7B model using WikiText-2 dataset.}
    \label{fig:std}
    \vspace{-1.5em} 
\end{figure}

\section{\ourmethod \space (\ourmethodabbr)}
Our initial investigation and prior literature confirm that randomly corrupting any bit does not disrupt LLM performance. To perform BFA without the knowledge of gradient and domain-specific data, an attacker has to develop a search algorithm that can identify specific regions of LLM that are vulnerable to bit corruption and, at the same time, independent of its application domain/data.  To develop this search algorithm, an attacker has to answer two questions: i) Which layer contains the vulnerable bit? \& ii) What is the index of the vulnerable bit at that layer?  

We propose the first \ourmethod \space (\ourmethodabbr), which can identify specific regions inside an LLM that are susceptible to bit flips, leveraging two vulnerability indices: The first component narrows the search to one particular layer by identifying the most sensitive layer in the model to bit corruption, which we label the Layer Vulnerability Index (LVI). Once the sensitive layer is selected, we propose a second metric, the Weight Vulnerability Index (WVI), to identify a vulnerable weight set. Neither of these indices requires any knowledge about the victim and can be computed using a publicly available dataset. Once the sequence of vulnerable weight indexes is identified for a specific LLM architecture, the attacker can transfer the locations of a bit-flips to any victim dataset or domain of application regardless of generative or classification task, without knowing the victim's application.

\subsection{Layer Vulnerability Index (LVI)}
\label{sensrtive_layer}

The first step of \ourattack \space is to identify the vulnerable layer index. To develop this identifier, we leverage a particular LLM property: outlier activation. In LLM, activations with extreme values are referred to as outlier activations~\cite{dettmers2022gpt3, sun2024massive} that exert a significant influence on the model’s behavior. The outlier activation can be measured by quantifying the shift in the standard deviation of activations between the input and output of each layer:
\begin{equation}
\Delta \sigma_\ell = |\sigma(h_\ell) - \sigma(h_{\ell-1})|,
\end{equation} \label{eq:LVI}

where a larger $\Delta \sigma_\ell$ indicates that the layer ($m$) significantly skews the distribution of activations; thus $m$ is the most vulnerable layer. As shown in Figure \ref{fig:std}, the second layer exhibits the maximum deviation and therefore is the most sensitive layer:
\begin{equation}
{L}_{m} = max (\Delta \sigma_\ell)
\label{eq:LVI}
\end{equation}

The intuition behind choosing outlier activation as an indicator for detecting a vulnerable layer to bit-flip is that this metric measures the deviation in activation as it passes through a given LLM layer, indicating that changes in this layer's weights can cause a larger shift in the output distribution. As a result, any bit-flip in this layer will cause a greater shift in output distribution from a non-corrupt LLM. Since our method is dataset-independent, we can utilize any publicly available datasets to detect the vulnerable layer. Our experiments demonstrate that the identification of the vulnerable layer remains consistent, regardless of the dataset used to compute the LVI, thereby serving our data-independent goal.

\subsection{Weight Vulnerability Index (WVI)}
\label{W_metric}
Even after narrowing the search space to a single sensitive layer, the number of weights within that layer remains substantial—often in the order of millions. To select a subset of weights in the sensitive layer as vulnerable, we define a Weight Vulnerability Index (WVI). Leveraging the phenomenon of \textit{outlier activations}, we conclude that weights associated with neurons exhibiting high activation magnitudes should be more susceptible to perturbation through bit-flipping. This hypothesis aligns with findings in pruning literature, where highly activated neurons are shown to have a disproportionate impact on model performance \cite{sun2024simpleeffectivepruningapproach}. Hence, we propose using the joint norm of the weights and activations to quantify the effect at the output neuron as our weight-sensitivity measurement.

Consider a linear layer with weights $W \in \mathbb{R}^{C_{\text{out}} \times C_{\text{in}}}$.
In LLMs, this layer receives input activations $A \in \mathbb{R}^{(B \times L) \times C_{\text{in}}}$,
where $B$ and $L$ denote the batch size and sequence length, respectively. For each weight $W_{ij}$, its importance score $I_{ij}$ is computed as the product of the absolute weight magnitude and the $\ell_2$-norm of the corresponding input feature vector:
\vspace{-0.2em}
\begin{equation}
{I}_{ij} = |W_{ij}| \cdot \|A_j\|_2
\label{eq:importance_score}
\end{equation}

The resulting scores $I_{ij}$ being high indicates that these vulnerable weights within the vulnerable layer will have a larger impact on the model's output behavior.

\subsection{Putting the Attack Together}
\label{flipping}
As shown in Figure\ref{fig:method}, the proposed GDF-BFA first feeds a sample of publicly available data to compute model outputs. Using the activation data, the attacker computes $\Delta \sigma$ in Eq.~\ref{eq:LVI} for every layer and picks the layer with the highest $\Delta \sigma$ value. Next, inside the vulnerable layer, the attacker computes the WVI metric in Eq.~\ref{eq:importance_score} to identify the vulnerable weight bit and flip it to achieve the desired attack goals. However, the WVI metric indicates a model's vulnerable weight index; it may not be the optimal or the only vulnerable weight index in the sensitive layer. Hence, we propose to select the top-\(k\) highest candidate weights in the sensitive layer instead:
\vspace{-0.2em}
\begin{equation}
\mathcal{W}_{\text{top-}k} = \operatorname{Top-K}(\mathcal{W}_{l}, I_{l}, k),
\label{eq:topk_selection}
\end{equation}
where $\mathcal{W}_{l}$ denotes the weights in sensitive layer $l$, $I_{l}$ represents their corresponding importance scores, and $\operatorname{Top-K}(\cdot)$ returns the $k$ weights with the highest WVI values. Each candidate is then evaluated by performing a single bit-flip independently and assessing the attacker's objective. We always flip the most significant bit (MSB) because its gradient will be higher and have the higher impact on the model's output.  The first weight whose MSB flip best achieves the attacker's objectives is selected for permanent bit flipping.

\begin{figure*}[t]  
    \centering
    \includegraphics[width=\textwidth,]{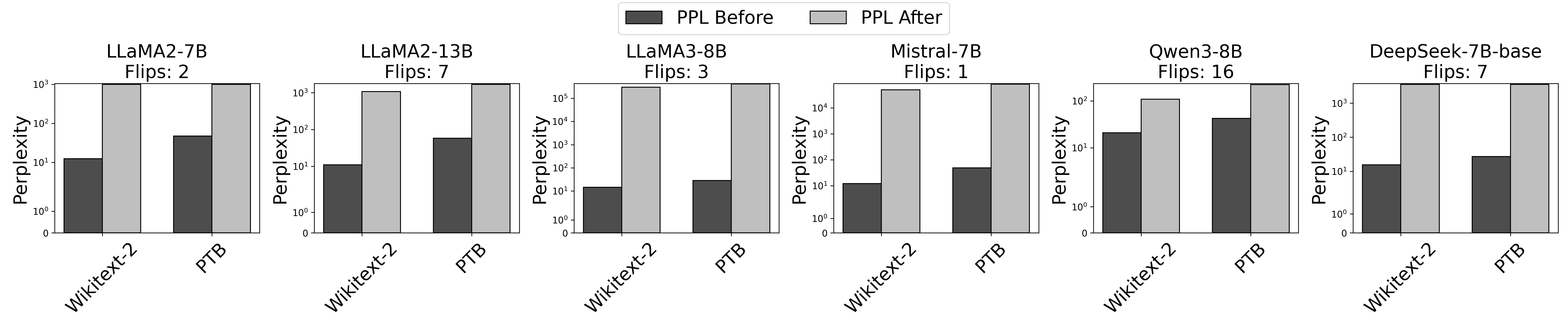}

    \caption{\textit{INT-8 Results on Generative Tasks}: PPL of \textbf{8-bit} models before and after \ourattack \space on WikiText-2 and PTB. \textit{A single implementation of \ourattack \space using ARC-Easy is sufficient to identify vulnerable weight bit locations; flipping these location collapse models performance on generative tasks by inducing PPL beyond 100.} \ourattack\ needs at most 16 bit flips to compromise the generative tasks of 8-bit LLMs.} 
    \label{fig:int8-gen-results}
    
\end{figure*}

\begin{figure*}[t]  
    \centering
    \includegraphics[width=0.93\textwidth,]{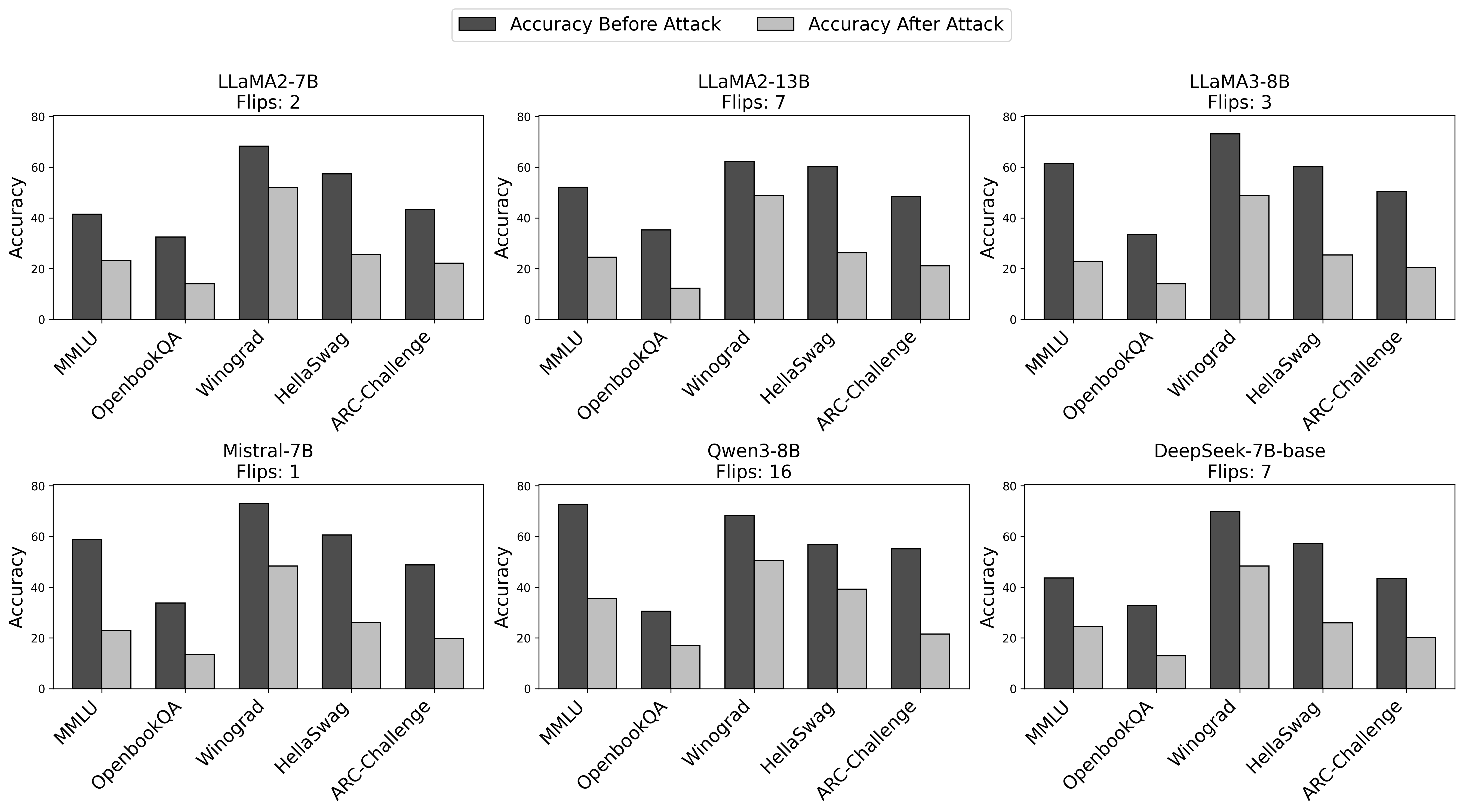}
    \vspace{-1em}
    \caption{\textit{INT-8 Results on Reasoning Tasks}: Accuracy (\%) of \textbf{8-bit} models before and after \ourattack \space across five reasoning and commonsense datasets. \textit{Flipping identified sequence of locations by \ourattack \space using ARC-Easy causes near-random performance ($\approx$25\% for four-class and $\approx$50\% for two-class tasks).} \ourattack\ needs as low as one bit flip to compromise the 8-bit LLMs.}
    \label{fig:int8-class-results}
    
\end{figure*}

\section{Evaluation}
\subsection{Experimental Setup}
\subsubsection{\textbf{Models and Datasets}}
We evaluate our attack on six open-source LLMs that cover a range of architectures and parameter sizes. Specifically, we include LLaMA 2-7B and LLaMA 2-13B \cite{touvron2023llama2openfoundation}, LLaMA 3-8B \cite{grattafiori2024llama3herdmodels}, Mistral 0.3-7B \cite{jiang2023mistral7b}, Qwen3-8B \cite{yang2025qwen3technicalreport}, and Deepseek-llm-7B-base \cite{deepseek-llm}. In addition, we evaluate these models under two precision settings: INT8 and INT4. For quantized evaluation, we use the BitsAndBytes library \cite{dettmers2022llmint88bitmatrixmultiplication}, which supports efficient 8-bit and 4-bit matrix multiplication with minimal degradation in model accuracy. 

To evaluate the impact of our attack, we used language modeling and reasoning benchmarks. For language modeling/generation, we use WikiText-2 \cite{merity2016pointersentinelmixturemodels} and Penn Treebank (PTB) \cite{marcus-etal-1993-building}, which measure the perplexity and token-level coherence of the generated text. To assess reasoning and commonsense understanding, we adopt the Massive Multitask Language Understanding (MMLU) benchmark \cite{gema2025mmlu} and the CommonsenseQA (CMQA) benchmark suite. The MMLU benchmark evaluates general knowledge and reasoning across 57 subjects. The CMQA suite consists of diverse commonsense reasoning tasks, we include the following:  HellaSwag \cite{zellers2019hellaswag}, WinoGrande \cite{sakaguchi2020winogrande}, ARC-Easy \cite{clark2018think}, ARC-Challenge \cite{clark2018think}, and OpenBookQA \cite{mihaylov2018can}. Both MMLU and CMQA consist of multiple-choice questions, and models must select the most likely answer.
All evaluations are performed using EleutherAI LM Harness \cite{eval-harness}.

\subsubsection{\textbf{Evaluation Metrics and Attack Settings}}\label{sec:eval}
To assess the impact of our attack on \textit{language modeling}, we use \textit{perplexity (PPL)}, which measures a model’s ability to generate coherent and contextually appropriate text. Lower PPL indicates stronger generative performance. For \textit{reasoning and commonsense benchmarks assessment}, we use \textit{accuracy}, defined as the proportion of correctly answered questions. 

For the \textbf{untargeted attack setting}, the objective of \ourattack\ is to increase the model’s PPL beyond 100, a threshold at which the model effectively fails to produce coherent text. In addition, the goal of the attack in this setting is to degrade the reasoning and commonsense benchmarks accuracy to the level of random. For datasets with four answer choices (MMLU \cite{gema2025mmlu}, HellaSwag \cite{zellers2019hellaswag}, ARC-Easy \cite{clark2018think}, ARC-Challenge \cite{clark2018think}, and OpenBookQA \cite{mihaylov2018can}), this corresponds to $\approx$25\% accuracy. For binary-choice tasks such as WinoGrande \cite{sakaguchi2020winogrande}, the random baseline is $\approx$50\%.

For the \textbf{targeted attack setting}, we apply the similar setting as prior adversarial weight attack described in \cite{rakin2021deep}. To evaluate performance under this setting, we report the \textit{Attack Success Rate (ASR)},  which is defined as the percentage of target class samples being incorrectly classified. In this setting, we target questions associated with a specific answer in a multiple-choice task where option \textit{C} is the target class for our experiments.

\subsection{Experimental Results and Analysis}
In all subsequent experiments, we use\textit{ only one dataset}, which is the publicly available ARC-Easy dataset \cite{clark2018think}, to identify the vulnerable components of the model. Specifically, ARC-Easy dataset is used to compute the \textit{LVI} metric, which determines the sensitive layer. It is also employed to evaluate the \textit{WVI} metric, which identifies vulnerable weights within the selected layer. In section~\ref{sec:data_indep}, we show that using different public datasets will result in identifying the same weight indexes. Consequently, all reported results are independent of gradient information and do not require access to training/testing data. We define attack success as the point at which the model’s accuracy on ARC-Easy decreases to $\approx$25\%, corresponding to random-guess performance in a four-option multiple-choice setting, or in the case where the accuracy does not decrease for $\geq 5$ iterations.

\begin{table*}[!t]
\centering
\caption{\textit{INT-4 Results:} PPL before/after the attack on WikiText-2 and PTB, and accuracy (\%) before/after the attack on the other five datasets. \textit{A single implementation of \ourattack\ using ARC-Easy identifies a sequence of vulnerable weight locations; flipping these locations collapses 4-bit quantized models across all domains.  Notably, \ourattack\ can compromise 4-bit quantized LLMs with as little as a single bit flip.}}
\resizebox{\textwidth}{!}{
\begin{tabular}{|c|c|c|c|c|c|c|c|c|}
\hline
\textbf{Model}      & \textbf{bit flips }& \textbf{Wikitext-2} & \textbf{PTB} & \textbf{MMLU} & \textbf{Openbookqa}  & \textbf{Winogrande} & \textbf{Hellaswag} & \textbf{Arc\_challenge} \\ \hline
Llama2-7B  & 1 & 13.80 / 665.14   & 53.751 / 909.14  & 39.51 / 23.5   & 32.6 / 14.6  &  68.58 / 48.5 & 57.02 / 28.4 & 42.66 /22.1 \\ \hline
Llama2-13B & 12 & 11.44 / 4074.48  & 58.11 / 2714.19  & 50.05 / 23.7& 35.2 / 15.0 &  71.1 / 51.5 & 59.41 / 26.9  & 46.92 / 21.7 \\ \hline
DeepSeek-7B-base  & 3 &16.91 / 236806.82 & 28.77 / 442413.39 & 41.4 / 24.7 & 31.0 / 14.2& 68.2 / 49.9& 56.1 / 25.7 & 43.4 / 21.8\\ \hline
Mistral-7B-v0.3  & 24  &  13.80 / 139.42&  56.33 / 367.32& 56.9 / 22.9&  32.8 / 15.8 & 72.6 / 53.5 &  59.8 / 30.6 &  46.9 / 20.9 \\ \hline
\end{tabular}
}
\label{tab:int4_results}
\end{table*}

\subsubsection{\textbf{Untargeted Attack in Quantized Models (8-Bits and 4-Bits).}}

Since quantized models are often assumed to be more robust to bit-level perturbations, we assess the effectiveness of \ourattack \space under two quantization settings: INT-8 (Figures~\ref{fig:int8-gen-results} and~\ref{fig:int8-class-results}) and INT-4 (Table~\ref{tab:int4_results}). Figure~\ref{fig:int8-gen-results} shows that executing \ourmethodabbr \space \textbf{\textit{only once}} on a single public dataset (ARC-Easy) to identify vulnerable bit locations is sufficient to severely degrade language modeling capability. Flipping the identified locations results in more than a \textbf{500$\times$ increase} in perplexity on WikiText-2 and a \textbf{20$\times$ increase} on PTB, demonstrating a substantial collapse in generative performance. Furthermore, as illustrated in Figure~\ref{fig:int8-class-results}, performance on reasoning and commonsense benchmarks drops to near-random accuracy, which indicate that the attacked models can no longer reliably select correct answers across diverse domains. These results confirm that the identified bits for flipping capture architecture level weaknesses rather than dataset-specific, and the flipping affect therefore transfers consistently across \textit{all downstream tasks}. The efficiency of \ourattack \space also extends to lower precision formats. As shown in Table~\ref{tab:int4_results}, INT-4 quantized models experience comparable performance collapse in both classification and generative evaluations, demonstrating the efficiency of \ourattack \space in lower precision quantized models.
Overall, these results show that our gray-box attack can identify a small set of vulnerable bits using only model architecture knowledge and publicly available data to compromise quantized LLMs across a wide range of downstream tasks.

\begin{tcolorbox}[
colback=gray!8,
colframe=black,
boxrule=0.6pt,
arc=2pt,
left=6pt,
right=6pt,
top=6pt,
bottom=6pt
]

\textbf{Observation 1: Efficient Gray-box BFA on LLMs.}
As shown in Figures~\ref{fig:int8-gen-results} and~\ref{fig:int8-class-results} and Table~\ref{tab:int4_results}, \ourattack\ identifies vulnerable weight locations using only architectural knowledge of the model, without requiring access to task-specific data, gradients, or domain information. This demonstrates a practical gray-box threat model with minimal attacker assumptions.

\medskip

\textbf{Observation 2: Effectiveness Across Quantization Levels.}
\ourattack\ remains effective under multiple precision settings, including INT-8 and INT-4 quantized models, achieving substantial performance degradation with only a small number of bit flips.

\medskip

\textbf{Observation 3: Strong Transferability Across Tasks.}
Once vulnerable bit locations are identified, flipping the same locations consistently degrades model performance across any downstream tasks. As demonstrated in Figures~\ref{fig:int8-gen-results} and~\ref{fig:int8-class-results} and Table~\ref{tab:int4_results}, identical flips compromise both generative evaluation and multiple commonsense and reasoning benchmarks without additional optimization.

\end{tcolorbox}

\subsubsection{\textbf{Targeted Attack.}}
Conventional targeted attacks typically require access to both training data and gradient information to steer the model toward the desired behavior. In contrast, \ourmethodabbr \space is explicitly designed to operate without access to gradients or data, making it considerably more challenging to achieve a fully targeted objective. However, if the threat model is slightly relaxed to allow the attacker access to the test set, targeted manipulation becomes feasible. This relaxation is a natural extension for targeted attack, since without knowing any knowledge of the data it is impossible to target a class. Nevertheless, the attack still remains gray-box since the weight and gradient knowledge is still restricted. 

In this setting, the attacker can influence model predictions to induce targeted behavior without relying on gradient or training data information. The available test set works as guide to the targeted miss-classification. The idea is that we still utilized LVI and WVI to identify the vulnerable bit index, but only select the candidate among the rank that directly impacts the classification of the target class C. As shown in Table~\ref{tab:targeted_attack}, \ourattack \space successfully achieved beyond 90\% ASR in three 8-bit models with the best performance achieved in LLaMA2-7B (100 \% ASR). The result is expected since targeted attack is more challenging in general, specially, without any gradient guidance.

\begin{tcolorbox}[
colback=gray!8,
colframe=black,
boxrule=0.6pt,
arc=2pt,
left=6pt,
right=6pt,
top=6pt,
bottom=6pt
]

\textbf{Observation 4: Extension to Targeted Attacks.}
When limited access to evaluation data is available, the gray-box (no access to weights or gradients) attack can be extended to induce targeted behavior, as demonstrated in Table~\ref{tab:targeted_attack}, where the test data is used to guide model's targeted behavior instead of gradient.

\end{tcolorbox}

\subsubsection{\textbf{Comparison with SOTA}}
In general, all the prior attacks on LLM using Bit-Flip are equally effective achieving adversative goals within the order of tens of bit flips across most architecture and dataset. However, our key contribution lies in relaxing the attack threat model in achieving the same objective.
As summarized in Table~\ref{tab:compare_SOTA}, existing BFAs on LLMs typically operate under white-box assumptions, requiring access to model weights, gradients, and task-specific datasets during optimization. In contrast, \ourattack\ aims to identify vulnerable bits under a gray-box setting, where the attacker knows only the model architecture and has no access to victim data or gradients.

Furthermore, prior BFAs incur high memory costs that scale significantly with model size, largely due to gradient-based optimization. In contrast, \ourattack\ requires nearly half the memory cost of previous BFAs by performing a fully gradient-free vulnerable-bit search.

\begin{table}[t]
\caption{\textit{Targeted Results:} Before-ASR and After-ASR refer to the ASR before and after the flipping on Arc-easy dataset, with target choice \textit{C}.}
\resizebox{0.9\columnwidth}{!}{
\begin{tabular}{|c|c|c|c|c|c|}
\hline
Model & 
\begin{tabular}[c]{@{}c@{}}bit-flips\end{tabular} & 
\begin{tabular}[c]{@{}c@{}}Before-ASR\end{tabular}  & 
\begin{tabular}[c]{@{}c@{}}After-ASR\end{tabular} \\ \hline
LLaMA2-7B      & 2  & 25.3  & 100.0 \\ \hline
Llama3-8B      & 10  &  26.05 &  90.14 \\ \hline
Mistral-7B-v0.3  &  4  &  16.9 &   90.14   \\ \hline
\end{tabular}}

\label{tab:targeted_attack}

\end{table}

\begin{table}[h]
\centering
\caption{\textbf{Comparison with SOTA BFAs on LLMs}: Threat model, victim data access, and memory cost. Memory cost is estimated for LLaMA-3-8B with batch size 16 and sequence length 1024.}
\resizebox{\columnwidth}{!}{
\begin{tabular}{|c|c|c|c|}
\hline
\textbf{Bit Flip Attack} & \textbf{Threat Model} & \textbf{Victim Data} & \textbf{Memory Cost} \\ \hline

GenBFA \cite{das2024genbfa}
& White-box & $\checkmark$ & $\approx$68 GB
\\ \hline

SBFA \cite{guo2025sbfa}
& White-box & $\checkmark$ & $\approx$68 GB
\\ \hline

SilentStriker \cite{xu2025silentstriker}
& White-box & $\checkmark$ & $\approx$68 GB
\\ \hline

FlipLLM \cite{khalil2025flipllm}
& White-box & $\checkmark$ & $\approx$68 GB
\\ \hline

\textbf{\ourattack\ (Ours)}
& \textbf{Gray-box}
& \textbf{\xmark}
& \textbf{$\approx$34 GB}
\\ \hline

\end{tabular}}
\label{tab:compare_SOTA}
\vspace{-1em}
\end{table}

\begin{figure*}
    \centering
    \includegraphics[width=\linewidth]{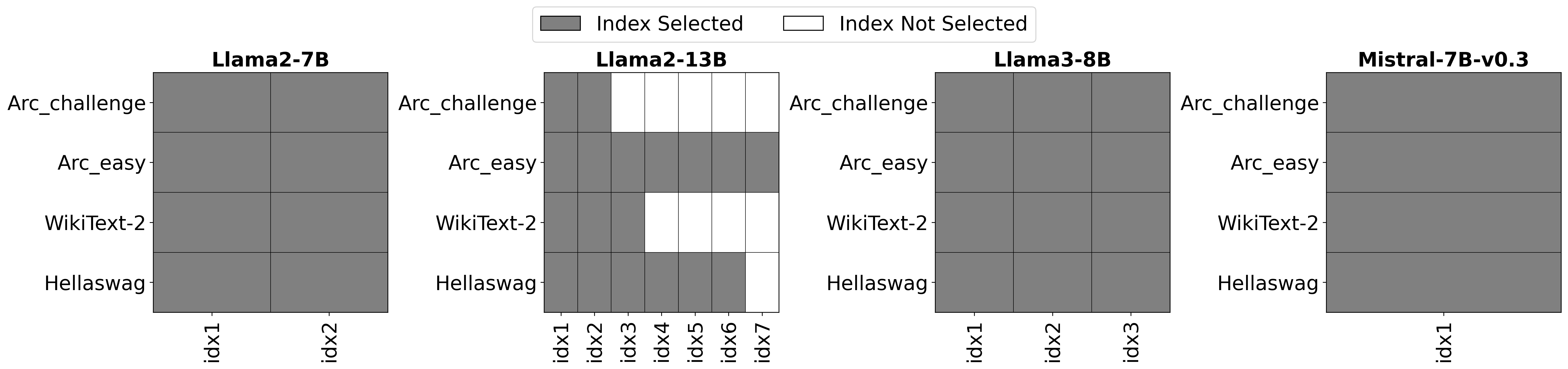}
    \caption{\textit{Selected vulnerable weight indices (WVI) identified using different datasets.} This figure demonstrates that, even when different publicly available datasets are used, the identified vulnerable weights remain largely consistent. Although LLaMA2‑13B selects additional weights in some cases, the most critical indices—particularly the first two selected weights—remain unchanged across datasets.}
    \label{fig:idx-selected}
    \vspace{-1em}
\end{figure*}

\begin{tcolorbox}[
colback=gray!8,
colframe=black,
boxrule=0.6pt,
arc=2pt,
left=6pt,
right=6pt,
top=6pt,
bottom=6pt
]

\textbf{Observation 5: Gradient-based search algorithms are inefficient for LLMs and require stronger attack privileges than \ourattack.}
As LLM sizes grow, gradient-based vulnerable-bit search becomes challenging to execute with low resources. Moreover, accessing victim datasets is increasingly challenging due to privacy constraints.

\end{tcolorbox}

\subsection{Insights and Analysis}
\subsubsection{\textbf{Selecting Vulnerable Weights with Different Public Datasets.}} \label{sec:data_indep}
To further demonstrate the efficiency of our method \ourmethodabbr, we show that the vulnerable weights identified for flipping remain largely consistent across different publicly available datasets. As illustrated in Figure~\ref{fig:idx-selected}, using four open-source datasets— WikiText-2 \cite{merity2016pointersentinelmixturemodels}, HellaSwag \cite{zellers2019hellaswag}, ARC-Easy \cite{clark2018think}, and ARC-Challenge \cite{clark2018think}—\ourmethodabbr\ consistently selects nearly the same set of vulnerable weights for the same model architecture. This consistency highlights that these weights are inherently critical for model performance, and These vulnerable indexes exist within the foundation models. Occasionally, \ourmethodabbr\ selects additional vulnerable weights, which may vary depending on target architecture and dataset.

\begin{tcolorbox}[
colback=gray!8,
colframe=black,
boxrule=0.6pt,
arc=2pt,
left=6pt,
right=6pt,
top=6pt,
bottom=6pt
]

\textbf{Observation 6: Vulnerable bit locations are largely data-independent in LLMs.}
As shown in Figure~\ref{fig:idx-selected}, vulnerable locations remain highly consistent across different datasets used for selection.

\end{tcolorbox}

\subsubsection{\textbf{Future Exploration of LLM Defense Mechanisms}}

Our attack and the proposed vulnerable-bit search algorithm reveal an important property of LLMs: \emph{each model architecture contains inherently vulnerable weight locations whose perturbation can degrade performance across any downstream tasks.} This observation suggests that model vulnerability is not purely task-dependent, but instead rooted in structural properties of the model itself. 

Leveraging this insight opens new directions for developing defense mechanisms against attacks. For example, as illustrated in Figure~\ref{fig:idx-selected}, safeguarding the vulnerable weights identified by GDF-BFA in Mistral-7B-v0.3 leads to a progressive increase in the number of bit flips required to compromise the model. As more critical weights are protected during each defense round, the attack complexity correspondingly increases. 

These findings suggest that studying this observed property of LLM can lead to  future defenses against weight-level attacks on LLMs.

\begin{tcolorbox}[
colback=gray!8,
colframe=black,
boxrule=0.6pt,
arc=2pt,
left=6pt,
right=6pt,
top=6pt,
bottom=6pt
]

\textbf{Takeaway: Implications for Future Defenses.}
Our findings reveal architecture-level vulnerabilities that are largely task-independent, highlighting new directions for developing robust defenses against attacks on LLMs.

\end{tcolorbox}

\begin{figure}[t]
    \includegraphics[width=0.8\columnwidth]{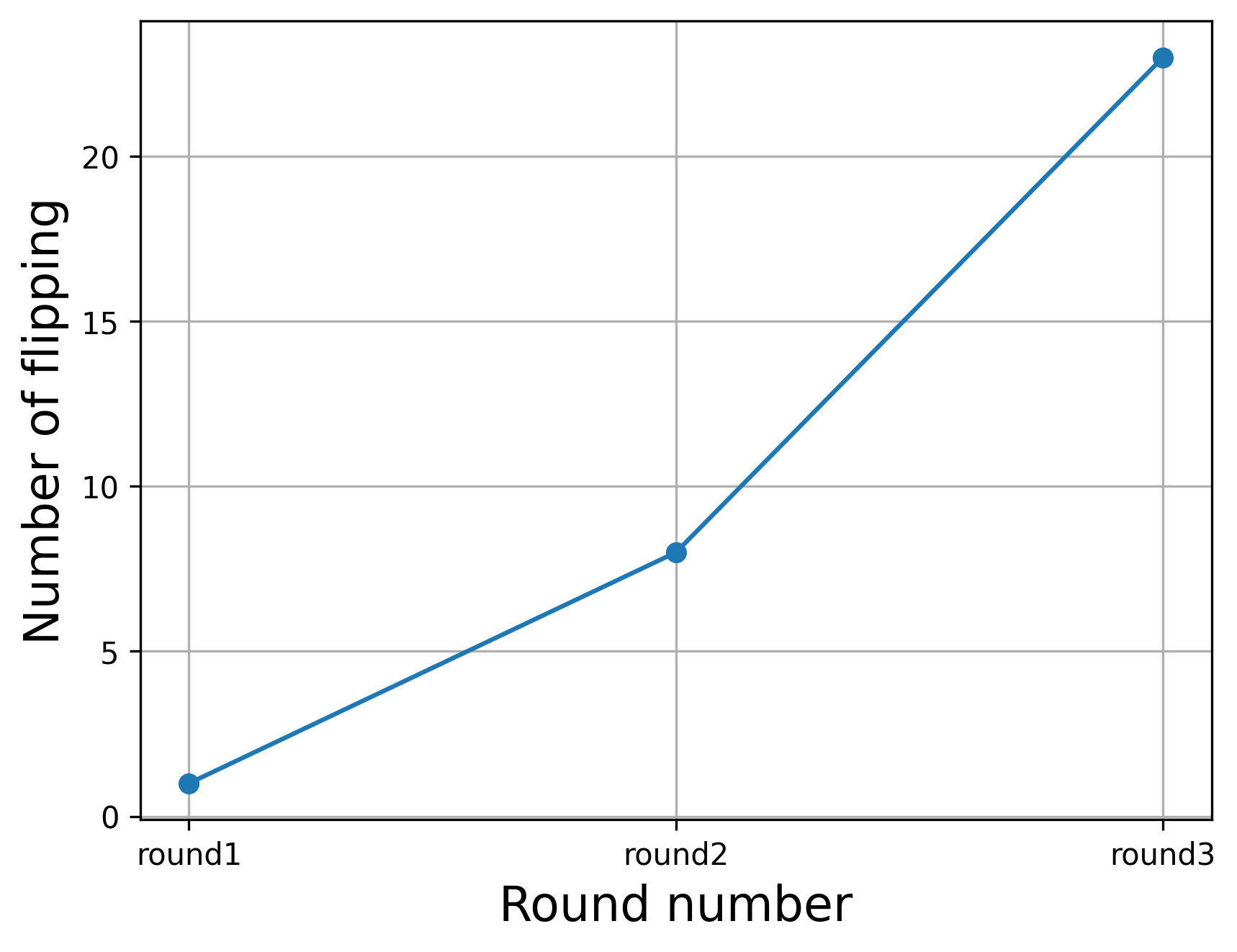} 
    \caption{Vulnerable weights identified by GDF-BFA are progressively protected (on Mistral-7B-v0.3 model) over three rounds: \textit{First round:} no weights protected. \textit{ Second round:} weights identified in first round protected;\textit{ Third Round:} weights identified in first and second rounds protected.}
    \label{fig:mistral-rounds}
    \vspace{-1em}
\end{figure}

\vspace{-1em}
\section{Conclusion}
Despite the remarkable advancements of LLMs in recent years, their vulnerability to bit-flip attacks remains a critical and largely underexplored security concern. Existing studies primarily operate under white-box assumptions and rely on gradient-based optimization to identify critical weights. For large-scale LLMs, such approaches are challenging, requiring substantial GPU resources as well as access to training or evaluation data.

In this work, we introduce the first gray-box bit-flip attack that requires neither gradient/weight information nor data access to locate vulnerable weights. Our method identifies sequences of weight locations whose single-bit corruption is sufficient to catastrophically degrade model performance, with the resulting damage consistently transferring across downstream tasks and evaluation domains. These findings reveal that LLM vulnerabilities can arise from architecture-level weaknesses rather than task-specific behaviors.

Overall, our results expose a previously unrecognized and severe security risk in modern LLMs and motivate the development of efficient defense mechanisms and robustness-aware model design for future deployments.


\end{document}